\documentclass[11pt]{article} 

\usepackage[utf8]{inputenc} 

\usepackage{geometry} 
\usepackage{graphicx} 

\usepackage{booktabs} 
\usepackage{array} 
\usepackage{paralist} 
\usepackage{verbatim} 
\usepackage{subfig} 
\usepackage{amsmath}
\usepackage{textcomp}
\usepackage{epstopdf}

\usepackage[nottoc,notlof,notlot]{tocbibind} 
\usepackage[titles,subfigure]{tocloft} 

\DeclareMathOperator{\Tr}{Tr}

\title{The exact solution for the free induction decay in a quasi-one-dimensional system in a multi-pulse NMR experiment}
\author{G.A.Bochkin$^{a,1}$, E.B.Fel'dman$^a$, S.G.Vasil'ev$^a$}
\newcommand{\todo}{}
\begin{document}
\maketitle
{\it $^a$Institute of Problems of Chemical Physics of Russian Academy of Sciences, Chernogolovka, Moscow region, 142432, Russia}

{\small $^1$E-mail address: bochkin.g@yandex.ru}
\vskip 1em

{\bf Abstract.} The exact solution for the free induction decay in a one-dimensional system in the multi-pulse experiment is obtained at both high and low temperatures in the approximation of nearest neighbor interactions. The experimental investigation is performed on a quasi-one-dimensional system of $^{19}$F nuclear spins in a single crystal of fluorapatite. The theoretical results are in a good agreement with the obtained experimental data.

\section{Introduction}\label{sec:intro}

The line shape of the nuclear magnetic resonance (NMR) spectrum and its Fourier transform, 
the free induction decay (FID), are important sources of information about structure and dynamics in condensed matter \cite{bloembergen,abragam,ernst}. At the same time, the problem \todo of the line shape in solids is intimately connected with the fundamental problem of statistical physics -- the establishment \todo of equilibrium in a many-particle system \cite{landafshitz_stat}.

Although the development of the theoretical methods for calculation of the FID began long ago \cite{abragam}, the complete solution of the problem is yet to be obtained and is the subject of ongoing research \cite{lundin2018,starkov2018}. More progress was achieved in one-dimensional systems where an expression of the $^{19}$F FID was calculated by exactly diagonalizing the Hamiltonian of the closed linear chain of five spins with nearest and next-nearest neighbor interactions \cite{engelsberg}.
 The effect of other nuclei was taken into account using a Gaussian broadening function of calculated multiplets \cite{engelsberg}.

The emergence of multi-pulse NMR spectroscopy \cite{haeberlen68} gave a new impulse \todo to the problem under consideration. The point is that the anisotropic dipole-dipole interactions (DDIs) are averaged out by the periodical multi-pulse sequence \cite{haeberlen68} and the remaining part of the DDI depends on the parameters of the sequence and the offset of the carrying frequency of the pulses from the Larmor frequency. For example, the remaining part of the DDI in the multi-pulse spin locking \cite{ivanov_ru} can change its sign resulting in a time reversal \cite{rhim70}. The non-secular two-quantum/two-spin Hamiltonian describing the behavior of the spin system on the preparation period of the multiple quantum NMR experiment \cite{baum}
 is the XY Hamiltonian \cite{dor} which can be exactly diagonalized in the approximation of nearest neighbor interactions. In the present letter, we use the multi-pulse sequence of the MQ NMR experiment \cite{baum} in order to investigate the NMR line shape at both high and low temperatures.

The letter is organized as follows. In Section~\ref{sec:fid} the FID in the multi-pulse NMR of a one-dimensional system with the pulse sequence used on the preparation period of the MQ NMR experiment at high temperatures is calculated analytically. The obtained experimental FID for a quasi-one-dimensional chain of $^{19}$F spin nuclei in a single crystal of fluorapatite is also presented. The analytical expression for that FID for low temperatures is given in Section~\ref{sec:analyt}. The results obtained are briefly summarized in Section~\ref{sec:summary}.

\section{The FID in the multi-pulse NMR experiment  of a one-dimensional system at high temperatures}\label{sec:fid}
\begin{figure}
\includegraphics{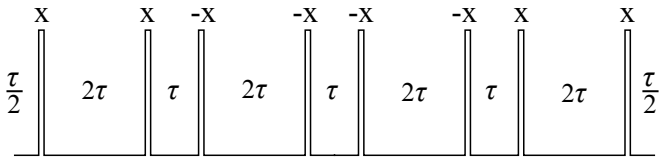}
\caption{Schematic representations of one period of the multi-pulse sequence used, consisting of resonance 
$\frac \pi 2$-pulses applied along the $x$-axis of the rotating reference frame; $\tau$, $2\tau$ are the delays between successive pulses}
\label{fig:pulseseq}
\end{figure}
Consider a one-dimensional system of $N$ nuclear spins ($I=\frac 12$) coupled by the DDI in the multi-pulse NMR experiment when it is irradiated by a periodic sequence of resonant 
$\frac \pi 2$-pulses applied along the $x$-axis of the rotating reference frame (RRF) \cite{goldman}. One period of the pulse sequence is presented in Fig.~\ref{fig:pulseseq}, where $\tau$ and $2\tau$ are the delays between successive pulses. The average Hamiltonian $H_{av}$ (the non-secular two-quantum/two-spin Hamiltonian \cite{baum, dor}) which describes the dynamics of the system, is, in the approximation of nearest neighbors, as follows:
\begin{equation}
H_{av}=  -\frac D2\sum\limits_{j=1}^{N-1}(I_j^+I_{j+1}^+ + I_j^-I_{j+1}^-) \label{hav},
\end{equation}
where $D$ is the DDI coupling constant of the nearest neighbors in the spin chain, and $I_j^+$,$I_j^-$ are the raising and lowering spin angular momentum operators of spin $j$. Since the axis of quantization here is the $z$-axis, we can conlude that $H_{av}$ is the so called XY-Hamiltionian \cite{dor}. This Hamiltonian can be diagonalized exactly \cite{dor}
and one can find the exact solution to the Liouville equation for the density matrix $\rho(t)$:

\begin{equation}
i\frac{d\rho(t)}{dt}=[H_{av},\rho(t)]\label{liouville}.
\end{equation}
We suppose that the initial condition for Eq.~\eqref{liouville} is the thermodynamic equilibrium state. The solution of Eq.~\eqref{liouville} can be written as \cite{dor}
\begin{equation}
  \rho(t) =   \rho_{0}(t) +  \rho_{2}(t) + \rho_{-2}(t) \label{rho},
\end{equation}
where
\begin{multline} \rho_0(t) = -\frac{2}{N+1} \sum\limits_k e^{-2i\varepsilon_k t} \Bigg\{ \sum\limits_{\genfrac{}{}{0pt}{}{l=1,3,\ldots, N-1}{l'=1,3,\ldots,N-1}}(-1)^{\frac{l+l'}2}2^{l+l'-2}\sin (kl) \sin(kl')\cdot\\ \cdot I_1^zI_2^z\ldots I_{l-1}^z I_l^+ I_1^zI_2^z\ldots I_{l'-1}^z I_{l'}^- + \sum\limits_{\genfrac{}{}{0pt}{}{l=2,4,\ldots, N}{l'=2,4,\ldots,N}}(-1)^{\frac{l+l'}2}2^{l+l'-2}\sin (kl) \sin(kl')\cdot\\ \cdot I_1^zI_2^z\ldots I_{l-1}^z I_l^- I_1^zI_2^z\ldots I_{l'-1}^z I_{l'}^+\Bigg\}\label{rho0},\end{multline}
\begin{multline}\rho_2(t) = -\frac{2}{N+1} \sum\limits_k e^{-2i\varepsilon_k t} \Bigg\{ \sum\limits_{\genfrac{}{}{0pt}{}{l=1,3,\ldots, N-1}{l'=2,4,\ldots,N}}(-1)^{\frac{l+l'+1}2}2^{l+l'-2}\sin (kl) \sin(kl')\cdot \\ \cdot I_1^zI_2^z\ldots I_{l-1}^z I_l^+ I_1^zI_2^z\ldots I_{l'-1}^z I_{l'}^+\Bigg \}\label{rho2},\end{multline}
\begin{multline}\rho_{-2}(t) = -\frac{2}{N+1} \sum\limits_k e^{2i\varepsilon_k t} \Bigg\{ \sum\limits_{\genfrac{}{}{0pt}{}{l=2,4,\ldots, N}{l'=1,3,\ldots,N-1}}(-1)^{\frac{l+l'+1}2}2^{l+l'-2}\sin (kl) \sin(kl')\cdot \\ \cdot I_1^zI_2^z\ldots I_{l-1}^z I_l^- I_1^zI_2^z\ldots I_{l'-1}^z I_{l'}^-\Bigg \} \label{rhomi2},\end{multline}

In Eqs.~(\ref{rho0}--\ref{rhomi2}) the fermionic energy $\varepsilon_k$ is \cite{dor}
\begin{equation}
\varepsilon_k=D\cos k,\qquad k=\frac{\pi n}{N+1},\qquad n=0,1,\ldots,N\label{energy},
\end{equation}
and $I_j^z$ is the projection operator of the spin angular momentum of the nucleus $j$ on the $z$ axis. When $N \gg 1$, it is possible to replace  the summation over $k$ in Eq.~\eqref{rho0} with integration. The $\rho_0(t)$ can be rewritten as
\begin{multline}\rho_0(t)=\sum\limits_{\genfrac{}{}{0pt}{}{l=1,3,\ldots, N-1}{l'=1,3,\ldots,N-1}}2^{l+l'-2}\left[J_{l-l'}(2Dt)+J_{l+l'}(2Dt)\right] I_1^zI_2^z\ldots I_{l-1}^z I_l^+ I_1^zI_2^z\ldots I_{l'-1}^z I_{l'}^- -\\- \sum\limits_{\genfrac{}{}{0pt}{}{l=2,4,\ldots, N}{l'=2,4,\ldots,N}}2^{l+l'-2} \left[J_{l-l'}(2Dt)-J_{l+l'}(2Dt)\right] I_1^zI_2^z\ldots I_{l-1}^z I_l^- I_1^zI_2^z\ldots I_{l'-1}^z I_{l'}^+\label{besselform},
\end{multline}
where $J_k$ is the Bessel function of the first kind of order $k$. One can find using Eq.~(\ref{liouville}) that
\begin{equation}\rho(t)=e^ {-iH_{av}t}\rho(0)e^{iH_{av}t}\label{evol},
\end{equation}
where the initial density matrix $\rho(0)$ can be considered to be equal to $I_z = \sum\limits_{i=1}^N I_j^z$ at high temperatures, up to a scale factor \cite{goldman}. The FID can be written as follows: \todo
\begin{equation}
G(t)=\Tr \left(\rho(t)I_z\right)\label{evolunreduced}.
\end{equation}
Taking into account that 
\begin{equation}
\Tr \left(\rho_2(t)I_z\right)=\Tr \left(\rho_{-2}(t)I_z\right)=0\label{zerotrace},
\end{equation}
and Eq.~(\ref{rho}), the expression for FID  \eqref{evolunreduced} becomes
\begin{equation}
G(t)=\Tr \left(\rho_0(t)I_z\right)\label{evolreduced}.
\end{equation}
Substituting Eq.~\eqref{besselform} into Eq.~\eqref{evolreduced} and simplifying, we get
\begin{equation}
G(t)=N2^{N-2}J_0(2Dt)-2^{N-2}\sum\limits_{l=1}^{\infty}(-1)^{l}J_{2l}(2Dt)\label{evolreduced2}.
\end{equation}
Now, using the relation for Bessel functions \cite{abramovitz}
\begin{equation}
J_0(z)+2\sum\limits_{l=1}^\infty(-1)^{l} J_{2l}(z)=\cos z \label{besselrel},
\end{equation}
one can obtain
\begin{equation}
G(t)=(2N+1)2^{N-3}J_0(2Dt) -2^{N-3}\cos 2Dt \label{fidn}.
\end{equation}
Normalizing $G(t)$ and taking the limit $N \to \infty$, we obtain the following formula for the FID:
\begin{equation}
\frac{G(t)}{G(0)}=J_0(2Dt) \label{fidfinal}.
\end{equation}
Taking into account that 
\begin{equation}
J_0(z) = 1 - \frac{z^2}4 + \ldots\label{besselseries},
\end{equation}
we can calculate the second moment of the line shape to be
\begin{equation}
M_2=-\left.\frac{d^2 J_0(2Dt)}{dt^2}\right|_{t=0}=2D^2.
\end{equation}
This corresponds to the local field acting on a spin, produced by its two nearest neighbors. We neglect the interactions of the remote spins in our calculations.

\begin{figure}
\includegraphics{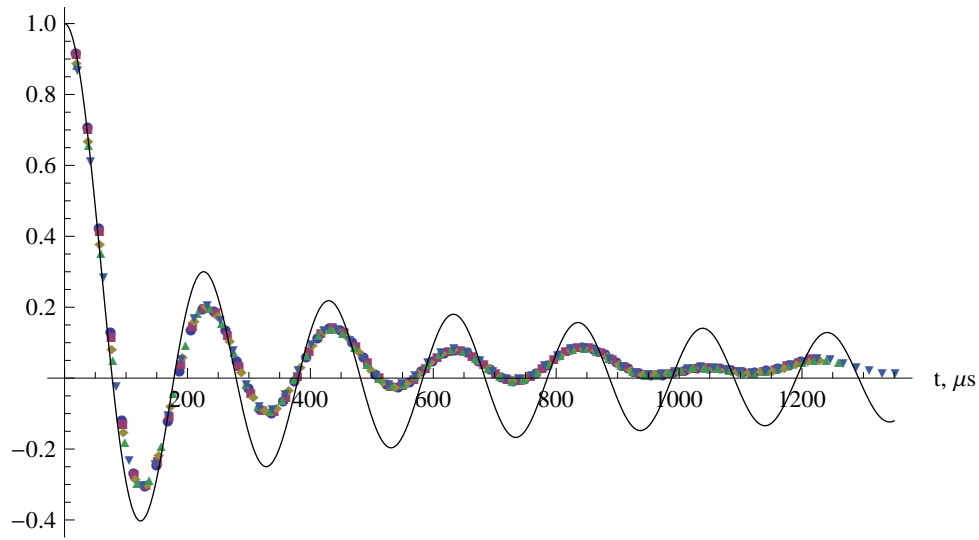}
\caption{The experimental and theoretical FIDs 
for a long open spin chain of $^{19}$F nuclei in a single crystal of fluorapatite; points are the experimental results, the line is the theoretical FID \eqref{fidfinal} with $D=15.5\cdot 10^3$~s$^{-1}$.}
\label{fig:fid}
\end{figure}

The experimental study of the FID was performed on a Bruker Avance III spectrometer with the static magnetic field of 9.4~T (the corresponding frequency on $^{19}$F nuclei is 376.6~MHz).  A probe with a solenoid coil for 2.5~mm o.d. NMR sample tubes was used. 
The time evolution of the FID was tracked by the repetition of the eight-pulse sequence of Fig.~\ref{fig:pulseseq} (from 1 to 64 times) followed by a 0.5~ms delay and a 90 degree detecting pulse. Five sets of such experiments with different $\tau$ = 0.8; 0.82; 0.84; 0.88 and 1 $\mu$s were performed.  The 90 degree pulse length was 0.76 $\mu$s. 8 scans were accumulated for each point. The dependence of the signal intensity on the length of the irradiation sequence is shown in Fig.~\ref{fig:fid}. One can see that the experimental FID agrees with the theoretical result \eqref{fidfinal}. A possible reason for the discrepancy is that the interactions of the remote $^{19}$F spins and heteronuclear interactions were neglected.

\section{The FID in the multi-pulse NMR experiment in a one-dimensional system at low temperatures}\label{sec:analyt}
We suppose that the system was initially in the thermodynamic equilibrium state:
\begin{equation}
\rho_{eq}=\frac{e^{\beta I_z}}{\Tr e^{\beta I_z}} \label{thermeq}.
\end{equation}
where $\beta=\frac{\hbar \omega_0}{kT}$, $\hbar$ and $k$ are the reduced Planck constant and Boltzmann constant respectively, $\omega_0$ is the Larmor frequency, $T$ is the temperature. The NMR line shape depends not only on $\rho_0(t)$ but also on $\rho_2(t)$ and $\rho_{-2}(t)$ (Eqs.~(\ref{rho2},\ref{rhomi2})) at low temperatures.

The density matrix $\rho(t)$ (Eq.~\eqref{rho}) can be expressed \cite{fl3} via the fermion field operators $a_k$, $a_k^+$ \cite{landafshitz_stat} 
 which obey the following anticommutation relations:
\begin{equation}
[a_j^+,a_k]_+=\delta_{jk},\qquad [a_j,a_k]_+=[a_j^+,a_k^+]_+=0\label{anticommut_rel}.
\end{equation}
The fermion representation of the density matrix $\rho(t)$ (Eq.~\eqref{rho}) \cite{FL2}
can be diagonalized using the Bogoliubov unitary transformations of the fermion operators \cite{mattis}:
\begin{eqnarray}
a_k=&u_k D_k & + v_k^*D_{-k}^+, \\a_{-k}=&-u_k D_{-k} & + v_k^*D_{k}^+\label{transf},
\end{eqnarray}
where $D_k$ and $D_k^+$ are the new representations of the fermion operators, and 
\begin{equation}
u_k=\sin(Dt \sin k),\qquad v_k=\cos(Dt \sin k)\label{uvdef}.
\end{equation}
Then the diagonal form of the density matrix $\rho(t)$ (Eq.~\eqref{rho}) is
\begin{equation}
\rho^{diag}(t)=\sum\limits_k \left(D_k^+ D_k -\frac 12 \right)\label{rhodiag}.
\end{equation}
The diagonal part of the operator $I_z = \frac N2 -\sum\limits_k a_k^+ a_k$ \cite{fl3} in the representation via $D_k$ can be written as follows:
\begin{equation}
I_z^{diag}= -\sum\limits_k \left(|u_k|^2-|v_k|^2\right) \left(D_k^+ D_k -\frac 12 \right)\label{izdiag}.
\end{equation}
In order to find the FID in the low-temperature case, one needs to calculate the correlator:
\begin{equation}
G(t)=\Tr\left(e^{-iH_{av}t}\rho_{eq}e^{iH_{av}t}I_z\right) =\frac 1Z \Tr\left(e^{\beta\rho(t)}I_z\right) = \frac 1Z\Tr\left(e^{\beta\sum\limits_k \left(D_k^+ D_k -\frac 12 \right)}I_z^{diag}\right)\label{correlator},
\end{equation}
where $\rho(t)=e^{-iH_{av}t}I_ze^{iH_{av}t}$ is the density matrix in the high-temperature case (discussed in Section~\ref{sec:fid}).
Using Eqs.~(\ref{uvdef}--\ref{izdiag}), one can obtain the final result 
\begin{equation}
G(t)= \frac N2\tanh\left(\frac{\hbar\omega_0}{2kT}\right)J_0(2Dt) \label{fidfinalunnorm},
\end{equation}
After normalization (such that $G(0)=1$), it is the same as the FID for high temperatures (see Eq.~\eqref{fidfinal}). 
The temperature dependence of the FID of Eq.~\eqref{fidfinalunnorm} is analogous to the temperature dependence of MQ NMR coherences at low temperatures \cite{fl3}.
\section{Conclusion}\label{sec:summary}
We developed the method for obtaining the FID in long open spin chains at both high and low temperatures in the approximation of DDIs of nearest neighbors. After normalization, the expression for the normalized FID at low temperatures is the same as the FID at high temperatures. The normalization constant depends on the temperature. The experimental FID, obtained for an open spin chain of $^{19}$F nuclei in a single crystal of fluorapatite, is in a good agreement with the theoretical results.
\section*{Acknowledgements}
This work was performed as part of the state task, state registration No.~0089-2019-0002. This work was performed using the equipment of the Center of Shared Analytical Facilities of IPCP RAS. This work is supported by the Russian Foundation for Basic Research, grant No.~19-32-80004. The authors acknowledge the support from Presidium of Russian Academy of Sciences, program No.~5 ``Photonic technologies in probing inhomogeneous media and biological objects''.

Declarations of interest: none.
\bibliographystyle{unsrt}
\bibliography{bibl}
\end{document}